\documentclass[11pt]{article}

\usepackage{amsmath}
\usepackage{graphicx}
\usepackage{amsfonts}
\usepackage{amssymb}
\usepackage{epsfig}
\usepackage{color}
\usepackage{psfrag}

\newcommand{\EQ}{\begin{equation}}
\newcommand{\EN}{\end{equation}}
\newcommand{\be}{\begin{equation}}
\newcommand{\ee}{\end{equation}}
\newcommand{\bea}{\begin{eqnarray}}
\newcommand{\eea}{\end{eqnarray}}

\def\th{\theta}

\begin{document} \setcounter{page}{0} 
\topmargin 0pt
\oddsidemargin 5mm
\renewcommand{\thefootnote}{\arabic{footnote}}
\newpage
\setcounter{page}{0}
\topmargin 0pt
\oddsidemargin 5mm
\renewcommand{\thefootnote}{\arabic{footnote}}
\newpage
\begin{titlepage}
\begin{flushright}
SISSA 40/2007/EP \\
DFTT 9/2007
\end{flushright}
\vspace{0.5cm}
\begin{center}
{\large {\bf Confinement in the $q$-state Potts field theory}}\\ 
\vspace{1.8cm}
{\large Gesualdo Delfino$^{a,b}$ and Paolo Grinza$^{c,d}$}\\
\vspace{0.5cm}
{\em ${}^a\,$International School for Advanced Studies (SISSA),}\\ 
{\em via Beirut 2-4, 34014 Trieste, Italy}\\
{\em ${}^b\,$Istituto Nazionale di Fisica Nucleare, sezione di Trieste, Italy}\\
\vspace{1mm}
{\em ${}^c\,$Dipartimento di Fisica Teorica dell'Universit\`a di Torino,}\\
{\em via P. Giuria 1, I-10125 Torno, Italy}\\ 
{\em ${}^d\,$Istituto Nazionale di Fisica Nucleare, sezione di Torino, Italy}\\
\end{center}
\vspace{1.2cm}

\renewcommand{\thefootnote}{\arabic{footnote}}
\setcounter{footnote}{0}

\begin{abstract}
\noindent
The $q$-state Potts field theory describes the universality class associated 
to the spontaneous breaking of the permutation symmetry of $q$ colors. In two
dimensions it is defined up to $q=4$ and exhibits duality and integrability 
away from critical temperature in absence of magnetic field. We show how, when
a magnetic field is switched on, it provides the simplest model of confinement
allowing for both mesons and baryons. Deconfined quarks (kinks) exist in a 
phase bounded by a first order transition on one side, and a second order 
transition on the other. The evolution of the mass spectrum with temperature 
and magnetic field is discussed.
\end{abstract}
\end{titlepage}

\newpage
\section{Introduction}
Confinement is that property of quantum field theory for which excitations which are genuine 
asymptotic particles in a region of coupling space become unobservable in isolation in another region, where they leave the place to new asymptotic particles  (mesons, baryons, ...) of which they can be seen as "constituents" (quarks).  The case of two-dimensional space-time provides the framework in which this, as other general properties of quantum field theory, can be studied in their simplest form. It is known that the gauge theory setting in which confinement is ordinarily discussed in four dimensions becomes somehow redundant in two dimensions. Indeed, due to the absence of transverse spatial dimensions, massless gauge fields do not carry particle degrees of freedom in $d=2$, so that an alternative description of the theory exists which relies on physical excitations only. For example, two-dimensional quantum electrodynamics can be exactly mapped onto the theory of a self-interacting neutral boson which makes quite transparent the presence of quark confinement  \cite{CJS,Coleman}. In absence of electromagnetic interaction the quarks correspond to the solitons interpolating between the vacua of a periodic bosonic potential. Quark interaction destroys the degeneracy of the bosonic vacua and removes the topologically charged excitations from the spectrum of asymptotic states. What remains is a spectrum of mesons originating from confinement of soliton-antisoliton pairs\footnote{We refer to the generic case in which the $\theta$-angle is not fine-tuned to the specific value which partially preserves vacuum degeneracy.}. 

This mechanism of confinement through breaking of degeneracy of discrete vacua is quite general in two dimensions and exhibits its most essential features in the case of a finite number of vacua originating from the spontaneous breaking of a discrete symmetry, with the kinks interpolating between these vacua playing the role of the quarks. Then it is not surprising that two-dimensional Ising field theory (i.e. the field theory describing the scaling limit of the two-dimensional Ising model) provides the simplest model of confinement (only two vacua). The associated mesonic spectrum was first studied in \cite{McW}. 

In theories, like Ising, with a one-component order parameter, the confined particles are made of an even number of quarks. Indeed, in this case the vacua are located along a line in order parameter space\footnote{This applies also to two-dimensional quantum electrodynamics, with the notion of one-dimensional order parameter referred to the bosonic version. See \cite{dsg} for a discussion of 
confinement in theories with more general bosonic potentials.}, so that the kink sequences starting from and going back to the true vacuum (the only ones generating bound states via confinement) consist of a number $2j$ of kinks. The lightest bound states 
are mesons (kink-antikink composites) corresponding to $j=1$. 

On the other hand, when the order parameter has more than one component we can find three vacua located on a plane in order parameter space. Now we can have a three-kink sequence making a loop through the three vacua. Such a sequence is confined into baryons, a finite number of which must be stable sufficiently close to the deconfining point. Indeed, if $m$ is the mass of the kink, the lightest baryons will have mass $\sim 3m$ and will not be able to  decay into two mesons with mass $\sim 2m$ each. 

In this paper we consider the simplest model of confinement allowing for baryons, i.e. the field theory describing the scaling limit of the two-dimensional $q$-state Potts model. The latter generalizes the Ising model to the case in which each site on the lattice can take $q$ different colors, and has an order parameter with $q-1$ components \cite{Potts,Wu}. In absence of magnetic field, the ferromagnetic $q$-state Potts model undergoes an ordering transition which is continuous up to a number of colors $q_c$, which in two dimensions equals 4 \cite{Baxter}. The scaling limit then can be taken up to this value of $q$ and produces a field theory in which a magnetic field confines the kinks and allows for baryons at $q=3,4$. 

In absence of the confining field, the quarks (kinks) behave as free neutral fermions in the Ising case ($q=2$), but become interacting for $q=3,4$. For $q=4$
the zero-field theory is equivalent to the sine-Gordon model at a specific 
value of the coupling ($\beta^2=2\pi$, see e.g. \cite{DG}), while for $q=3$ the
ultraviolet fixed point is non-trivial. In any case the confining interaction
is non-local with respect to the quarks. The possibility of a quantitative 
study in weak field comes from integrability of the $q$-state Potts field 
theory in zero-field \cite{CZ}. Hence, form factor perturbation theory 
\cite{nonint} allows to express mass corrections in terms of the matrix 
elements of the magnetic operator computed in \cite{DC}. 

Here, however, our 
main interest will be in a qualitative characterization of the evolution of the
mass spectrum for generic values of the temperature and of a magnetic field 
chosen to act on a single color. For $q=3,4$ this choice allows for an
extended phase on the parameter plane in which the quarks are deconfined. Such 
a phase is bounded by the confining (first order) transition on one side, and 
by a spontaneous breaking (second order) transition on the other. Outside 
this region, the spectrum of asymptotic particles is made of mesons 
(everywhere) and baryons (at least sufficiently close to the deconfining
transition).

The paper is organized as follows. In the next section we discuss the $q$-state
Potts model with our choice of magnetic field, starting with the lattice 
definition and then switching to the field theoretical description of the 
scaling limit. In section~3 we focus on the two-dimensional case and recall the
exact scattering solution for the low-temperature phase in zero-field, before
showing how this is related by duality to the scattering solution for the 
high-temperature phase. Section~4 is devoted to the weak field analysis, while
the spectrum evolution as a function of temperature and magnetic field is 
discussed in section~5. Few final remarks are collected in section~6.

\section{Potts model with magnetic field}
In this section we discuss the lattice definition of the $q$-state Potts
model and the field theoretical description of the scaling limit. Some 
remarks about $d>2$, and about $q>4$ in $d=2$ are included although they 
are not used in the rest of the paper.

\subsection{Lattice model}
The $q$-state Potts model \cite{Potts,Wu} is a generalization of the Ising
model in which each site variable $s(x)$ at site $x$ on the lattice can 
assume $q$ different values (colors).
In absence of magnetic field the interaction only distinguishes whether 
nearest neighbor sites have equal or different color, so that the 
Hamiltonian is invariant under the group $S_q$ of permutation of the colors.
If we add a magnetic field $H$ acting only on the sites with a specific color
(say $s=q$), the reduced Hamiltonian can be written as
\EQ
{\cal H}=-\frac{1}{T}\sum_{(x,y)}\delta_{s(x),s(y)}-H\sum_{x}\delta_{s(x),q}\,,
\label{lattice}
\EN
and is invariant under the group $S_{q-1}$ of permutations of the first 
$q-1$ colors (the first sum is over nearest neighbors).

In the ferromagnetic case at $H=0$, the $q$ configurations in which all the 
sites 
have the same color minimize the energy and the system exhibits spontaneous 
magnetisation for sufficiently low values of the temperature $T$. Above  
a critical temperature $T_c$ the thermal fluctuations become 
dominant and the system is in a disordered phase. If we introduce the variables
\EQ
\sigma_\alpha(x)=\delta_{s(x),\alpha}-\frac{1}{q}\,,
\hspace{1cm}\alpha=1,2,\ldots,q
\label{sigma}
\EN
satisfying the condition
\EQ
\sum_{\alpha=1}^q\sigma_\alpha(x)=0\,,
\label{constraint}
\EN
the expectation values $\langle\sigma_\alpha\rangle$ differ from zero only in 
the low-temperature phase and can be used as order parameters.

When the magnetic field is switched on with a positive value, the ground state
at $T=0$ is unique (all sites have color $q$), and there can be no phase 
transition as the temperature is increased. 

Different is the situation for $H<0$, $q>2$. As $H\to-\infty$ the
color $q$ becomes forbidden and a zero-field $(q-1)$-Potts model is obtained. 
Then the critical points at $H=0$ and $H=-\infty$ are 
the endpoints of a phase transition line which in the $T$-$H$ plane 
separates a low-temperature, spontaneously magnetized phase with $q-1$ 
degenerate ground states from a high-temperature, disordered phase (Fig.~1). 

The nature of the transition at $T=T_c(H)$, $H\leq 0$, depends on $q$ and on 
the dimensionality $d$. It is well known (see \cite{Wu}) that in the 
zero-field $q$-state Potts model there exists a value $q_c(d)$ (not 
necessarily integer\footnote{One can make sense of the Potts model
for non-integer values of $q$ through the mapping onto the random cluster model
\cite{KF}. Although we will have mainly in mind integer values, most of our 
discussion can be done treating $q$ as a continuous parameter, and this will be
understood in the following.}) such that the transition is continuous for 
$q\leq q_c$ and first order for $q>q_c$. Accordingly, three cases can be 
distinguished for the transition going from $C_q$ to $C_{q-1}$ in Fig.~1:

\vspace{.3cm}
\noindent
i) $q-1>q_c$. The transition is first order with $q$ phases coexisting 
along the transition line. Only at $C_q$ $q+1$ phases coexist.

\noindent
ii) $2<q\leq q_c$. The transition is continuous.

\noindent
iii) $q-1\leq q_c<q$. The correlation length is finite at $C_q$ and infinite
at $C_{q-1}$. The nature of the transition depends on $d$.
 
\vspace{.3cm}
The value $q_c(2)$ is exactly known to be 4 \cite{Baxter}, while $q_c(3)$ lies 
in between 2 and 3. Hence, in $d=2$ the transition induced by the field is 
continuous for $q=3,4$ and first order for $q>5$; in $d=3$ it is first
order for $q\geq 4$. The cases $q=5$ in $d=2$ and $q=3$ in $d=3$ are of the 
type iii) above and will be discussed in a moment.

\begin{figure}
\centerline{
\includegraphics[width=9cm]{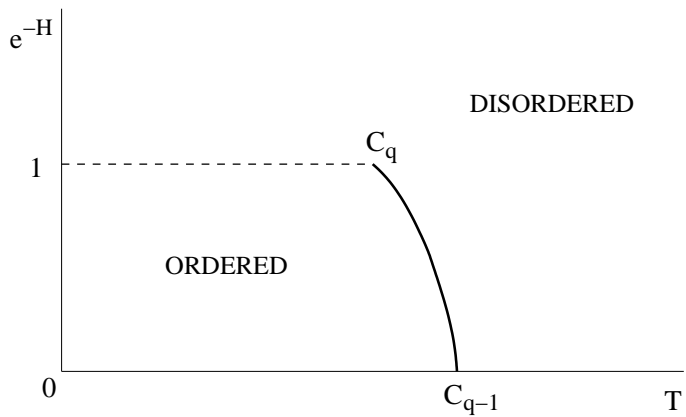}}
\caption{Phase diagram of the model (\ref{lattice}) for $q>2$. The values 
$H=0$ and $H=-\infty$ correspond to the zero-field $q$- and 
$(q-1)$-state Potts model, respectively. The ordered
phase possesses $q-1$ degenerate ground states and is separated from the 
disordered region by two phase transition lines (dashed and thick line). 
The dashed line is a first order transition line along which $q$ ground
states are degenerate. The nature of the transition along the thick line
depends on $q$ and $d$.
}
\end{figure}

\subsection{Field theory description}
A continuous, field theoretical description at scales much larger than the 
lattice spacing is possible at and around those points of the phase diagram 
where the correlation length diverges. For $q\leq q_c$, the transition point
at $H=0$ ($C_q$ in Fig.~1) corresponds in the scaling limit to a fixed point
of the renormalization group, i.e. to a conformal field theory with action
${\cal A}_{CFT}^{(q)}$. The scaling limit of (\ref{lattice}) around $C_q$ is 
described by the action
\EQ
{\cal A}={\cal A}_{CFT}^{(q)}-\tau\int d^dx\,\varepsilon(x)-
h\int d^dx\,\sigma_q(x)\,\,,
\label{scaling}
\EN
where $\varepsilon(x)$ is the leading $S_q$-invariant operator in 
${\cal A}_{CFT}^{(q)}$, and and $\sigma_q(x)$ is the leading 
$S_{q-1}$-preserving magnetic operator. If $X_\Phi^{(q)}$ denotes the scaling 
dimension of an operator 
$\Phi(x)$ at the $S_q$-invariant fixed point and $m$ a mass scale, 
the couplings $\tau$ and $h$ behave dimensionally as
\EQ
\tau\sim m^{d-X_\varepsilon^{(q)}}\,,\hspace{1.5cm}h\sim m^{d-X_\sigma^{(q)}}
\,,
\EN
and measure the deviation from critical temperature and the magnetic field,
respectively. 


For $2<q\leq q_c$ the transition along the line joining $C_q$ to $C_{q-1}$ 
in Fig.~1 is continuous and the scaling action (\ref{scaling}) with $\tau=0$, 
$h<0$ describes a massless flow from ${\cal A}_{CFT}^{(q)}$ to 
${\cal A}_{CFT}^{(q-1)}$. The scaling limit around the infrared fixed point
$C_{q-1}$ is described by the action
\EQ
{\cal A}_{IR}={\cal A}_{CFT}^{(q-1)}-\tilde{\tau}\int d^dx\,\varepsilon(x)+
\lambda\int d^dx\,\phi(x)+\ldots\,\,,
\label{ir}
\EN
where $\tilde{\tau}\sim m^{d-X_{\varepsilon}^{(q-1)}}$ is proportional to
$T-T_c$ and $\lambda\sim m^{d-X_{\phi}^{(q-1)}}$ is proportional to
$1/H$. All the operators in the r.h.s. of (\ref{ir}) are $S_{q-1}$-invariant:
$\varepsilon$ is relevant, while $\phi$ is the most relevant of 
the infinitely many irrelevant operators (dots) which specify the massless 
flow at $\tau=0$.

In $d=3$ the condition $2<q\leq q_c$ is not satisfied for integer values of 
$q$ and the transition at $H<0$ is first order for $q>3$. For $q=3$ the 
correlation length is finite at $C_q$ and infinite at $C_{q-1}$. Since the 
latter is an Ising critical point for which $\varepsilon$ is the only 
symmetry-preserving relevant operator, $C_2$ must be an infrared fixed point 
whose scaling region is described by the action (\ref{ir}). The ultraviolet
endpoint of the massless flow at $\tilde{\tau}=0$ must be a fixed point $P$ 
located on the transition line (Fig.~2a). The nature of the transition then
requires that $P$ is an Ising tricritical point\footnote{Since Ising 
tricriticality is described by a $\Phi^6$ Landau-Ginzburg potential for which 
$d=3$ is the upper critical dimension, $P$ is a Gaussian fixed point.}.

\begin{figure}
\centerline{
\includegraphics[width=14cm]{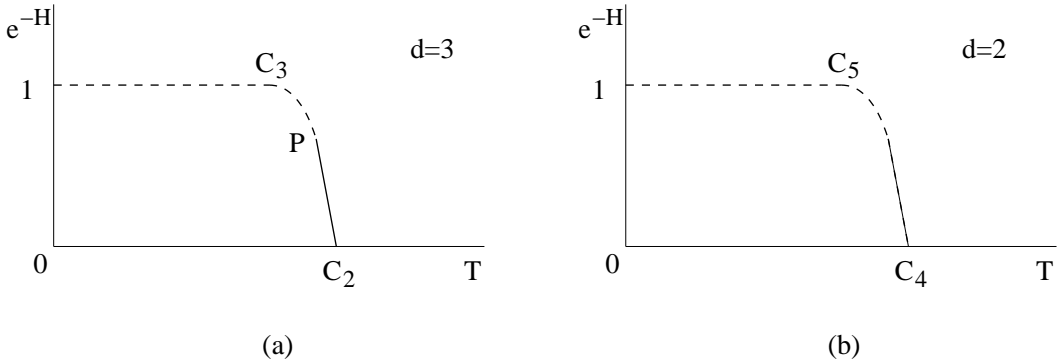}}
\caption{Phase diagrams of the model (\ref{lattice}) for $d=q=3$ (a), and
for $d=2$, $q=5$ (b). The correlation length is infinite along $PC_2$ and 
at $C_4$; $q$ phases coexist at generic points along the dashed lines, $q+1$
at the points $C_3$ and $C_5$.
}
\end{figure}

In $d=2$ the field theoretical description can rely on a number of exact 
results. Baxter solved the zero-field $q$-state Potts model at $T=T_c$ on the
square lattice and found that $q_c=4$ \cite{Baxter}. The scaling limit of 
the critical point up to $q_c$ was later identified \cite{DF} to correspond to 
the conformal field theory with central charge \cite{BPZ}
\EQ
c(q)=1-\frac{6}{t(t+1)}\,,
\label{c}
\EN
where the parameter $t$ is related to $q$ by the formula
\EQ
\sqrt{q}=2\sin\frac{\pi(t-1)}{2(t+1)}\,\,.
\EN
The scaling dimensions of the leading thermal and magnetic operators 
coincide with those of the operators $\phi_{2,1}$ and $\phi_{(t-1)/2,(t+1)/2}$
in the conformal theory, and read \cite{DF,Nienhuis}
\EQ
X_\varepsilon^{(q)}=\frac{1}{2}\left(1+\frac{3}{t}\right)\,,
\hspace{1.5cm}
X_\sigma^{(q)}=\frac{(t-1)(t+3)}{8t(t+1)}\,\,.
\label{xsigma}
\EN
The action (\ref{scaling}) with $d=2$ describes the scaling region around 
$C_q$ for $q\leq 4$ and is known to correspond to an integrable quantum field
theory for $h=0$ \cite{CZ} (see next section). For $\tau=0$, $h<0$ and 
$2<q\leq 4$ it describes a 
massless flow between an ultraviolet fixed point with central charge $c(q)$ 
and an infrared fixed point with central charge $c(q-1)$. Around this latter
fixed point we can use the action (\ref{ir}) with the irrelevant operator
$\phi$ identified with the operator $\phi_{3,1}$ of the conformal 
classification. Its scaling dimension at an $S_q$-invariant critical point
is 
\EQ
X_\phi^{(q)}=2\left(1+\frac{2}{t}\right)\,\,.
\EN
A field theory description around $C_{q-1}$ is still possible at $q=5$. 
Since $X_\phi^{(4)}=2$, $\phi$ is in this case a marginal operator
and the scaling region around $C_4$ is described by the action
\EQ
{\cal A}_{CFT}^{(4)}-\tilde{\tau}\int d^2x\,\varepsilon(x)+
\lambda\int d^2x\,\phi(x)\,\,.
\label{q4}
\EN
Since the two-dimensional 4-state Potts model does not admit a tricritical 
point\footnote{A phase diagram analogous to that of Fig.~2a, with $P$ 
corresponding to a $(q-1)$-Potts tricritical point, should be obtained for
$4<q<5$. $P$ tends to $C_q$ as $q\to 4$ and to $C_{q-1}$ as $q\to 5$.}
 \cite{NBRS,CNS}, $\phi$ acts as a marginally relevant perturbation and
the transition is first order (Fig.~2b). The action (\ref{q4}) is 
integrable also for $\tau=0$ \cite{dilute,q4} and can be used to describe 
exactly the transition close to $C_4$.

\section{Exact scaling solution in zero field}
In the two-dimensional case, to which we restrict our attention from now on, 
the action (\ref{scaling}) is integrable for $h=0$ and the solution 
can be found in the form of an exact, elastic and factorized $S$-matrix for the
relativistic particles of the associated $(1+1)$-dimensional theory \cite{ZZ}. 
The scattering theories above and below $T_c$ must describe two different 
physical situations and, at the same time, must reflect the existence of a
duality transformation \cite{Potts,Wu} relating the ordered and disordered 
phases.

\subsection{Ordered phase}
The $S$-matrix in the case of spontaneously broken symmetry ($\tau<0$) was
determined\footnote{The first exact $S$-matrix for $\phi_{2,1}$--perturbed 
conformal field theories was determined in \cite{Smirnov12}. The relation 
between this solution, which does not exploit $S_q$ invariance and relies on 
a different particle basis, and the one of \cite{CZ} is explained in \cite{FR}.
} in \cite{CZ}. The $q$ ferromagnetic ground states correspond in the 
field theory to degenerate vacua labelled by an index $\alpha=1,2,\ldots,q$. 
The elementary excitations are then provided by kinks\footnote{We parameterise 
on-shell momenta as $p^\mu=(m\cosh\theta,m\sinh\theta)$, $m$ being the mass
of the kink.} $K_{\alpha\beta}(\theta)$ interpolating between the vacua 
$\alpha$ and $\beta$ ($\alpha\neq\beta$). The space of asymptotic states 
consists of multi-kink configurations of the type 
$K_{\alpha_0\alpha_1}(\theta_1)K_{\alpha_1\alpha_2}(\theta_2)\ldots 
K_{\alpha_{n-1}\alpha_n}(\theta_n)$ ($\alpha_i\neq\alpha_{i+1}$) interpolating 
between the vacua $\alpha_0$ and $\alpha_n$. As a consequence of 
$S_q$-invariance, all the $n$-kink states fall into two topological sectors:
the neutral sector, corresponding to $\alpha_0=\alpha_n$, and the 
charged sector, corresponding to $\alpha_0\neq\alpha_n$.

Integrability implies that the scattering processes are completely elastic and
factorised into the product of two-kink interactions. An outgoing two-kink 
state can only differ from the ingoing one 
by the vacuum state between the kinks. Hence, the two-kink scattering can 
formally be described through the Faddeev-Zamolodchikov commutation relation
\EQ
K_{\alpha\gamma}(\theta_1)K_{\gamma\beta}(\theta_2)=\sum_{\delta\neq\alpha,
\beta}S_{\alpha\beta}^{\gamma\delta}
(\theta_{12})K_{\alpha\delta}(\theta_2)K_{\delta\beta}(\theta_1)\,,
\label{fz}
\EN
where $\theta_{12}\equiv\th_1-\theta_2$, and $S_{\alpha\beta}^{\gamma\delta}
(\theta_{12})$ 
denote the two-body scattering amplitudes (Fig.~3a). $S_q$ invariance reduces
to four the number of independent amplitudes, two for the charged and two
for the neutral topological sector
\bea
&& K_{\alpha\gamma}(\theta_1)K_{\gamma\beta}(\theta_2)=S_0(\theta_{12})
\sum_{\delta\neq\gamma}K_{\alpha\delta}(\theta_2)K_{\delta\beta}(\theta_1)+
S_1(\theta_{12})
K_{\alpha\gamma}(\theta_2)K_{\gamma\beta}(\th_1)\,,\hspace{.5cm}\alpha\neq\beta
\nonumber\\
&& K_{\alpha\gamma}(\theta_1)K_{\gamma\alpha}(\theta_2)=S_2(\theta_{12})
\sum_{\delta\neq\gamma}K_{\alpha\delta}(\theta_2)K_{\delta\alpha}(\theta_1)+
S_3(\theta_{12})K_{\alpha\gamma}(\theta_2)K_{\gamma\alpha}(\theta_1)\,\,.
\eea
Using the commutation relation (\ref{fz}) twice one obtains the unitarity
constraint
\EQ
\sum_{\varepsilon\neq\alpha,\beta} S_{\alpha\beta}^{\gamma\varepsilon}(\theta)
S_{\alpha\beta}^{\varepsilon\delta}(-\theta)=\delta^{\gamma\delta}\,,
\EN
which amounts to the set of equations
\bea
&& (q-3)S_0(\theta)S_0(-\theta)+S_1(\theta)S_1(-\theta)=1\,,
\label{uni1}\\
&& (q-4)S_0(\theta)S_0(-\theta)+S_0(\theta)S_1(-\theta)+S_1(\theta)S_0(-\theta)
=0\,,\\
&& (q-2)S_2(\theta)S_2(-\theta)+S_3(\theta)S_3(-\theta)=1\,,\\
&& (q-3)S_2(\theta)S_2(-\theta)+S_3(\theta)S_2(-\theta)+S_2(\theta)S_3(-\theta)
=0\,\,.
\label{uni4}
\eea
Crossing symmetry provides the relations 
\bea
&& S_0(\theta)=S_0(i\pi-\theta)\,\label{cross1}\\
&& S_1(\theta)=S_2(i\pi-\theta)\,\\
&& S_3(\theta)=S_3(i\pi-\theta)\,\,.\label{cross3}
\eea
Using these constraints together with the Yang-Baxter and bootstrap equations 
(that we do not reproduce here) the following expressions for the 
four elementary amplitudes were determined in Ref.\,\cite{CZ}
\bea
&& S_0(\theta)=\frac{\sinh\lambda\theta\,\sinh\lambda(\theta-i\pi)}
{\sinh\lambda\left(\theta-\frac{2\pi i}{3}\right)\,\sinh\lambda\left(\theta-
\frac{i\pi}{3}\right)}\,\Pi\left(\frac{\lambda\theta}{i\pi}\right)\,,
\label{s0}\\
&& S_1(\theta)=\frac{\sin\frac{2\pi\lambda}{3}\,\sinh\lambda(\theta-i\pi)}
{\sin\frac{\pi\lambda}{3}\,\sinh\lambda\left(\theta-\frac{2 i\pi}{3}\right)}\,
\Pi\left(\frac{\lambda\theta}{i\pi}\right)\,,\\
&& S_2(\theta)=\frac{\sin\frac{2\pi\lambda}{3}\,\sinh\lambda\theta}
{\sin\frac{\pi\lambda}{3}\,\sinh\lambda\left(\theta-\frac{i\pi}{3}\right)}\,
\Pi\left(\frac{\lambda\theta}{i\pi}\right)\,,\label{s2}\\
&& S_3(\theta)=\frac{\sin\lambda\pi}{\sin\frac{\pi\lambda}{3}}\,
\Pi\left(\frac{\lambda\theta}{i\pi}\right)\,,\label{s3}
\eea
where $\lambda$ is related to $q$ as
\EQ
\sqrt{q}=2\sin\frac{\pi\lambda}{3}\,,
\label{qlambda}
\EN
and
\bea
&&\Pi\left(\frac{\lambda\theta}{i\pi}\right)=
\frac{\sinh\lambda\left(\theta+i\frac{\pi}{3}\right)}{\sinh\lambda(\theta-
i\pi)}\,e^{{\cal A}(\theta)}\,,\\
&& {\cal A}(\theta)=\int_0^\infty\frac{dx}{x}\,
\frac{\sinh\frac{x}{2}\left(1-\frac{1}{\lambda}\right)-
\sinh\frac{x}{2}\left(\frac{1}{\lambda}-\frac{5}{3}\right)}
{\sinh\frac{x}{2\lambda}\cosh\frac{x}{2}}\,\sinh\frac{x\theta}{i\pi}\,\,.
\eea
The above solution is well defined for real values of $\lambda$, and one sees
that (\ref{qlambda}) implies $0\leq q\leq 4$. The pure scaling Potts model
in this range of $q$ is identified with the values of $\lambda$ in the range
going from\footnote{The range $3/2<\lambda<3$ corresponds to the thermal
perturbation of the tricritical $q$-state Potts model \cite{CZ}.} 0 to $3/2$.

The function $\Pi(\lambda\th/i\pi)$ is free of poles in the physical strip 
$\mbox{Im}\,\theta\in(0,\pi)$ for $q<3$ (i.e. $\lambda<1$).
Hence, in this range of $q$ the only poles of the scattering amplitudes in 
the physical strip are those located at $\theta=2i\pi/3$ and $\theta=i\pi/3$ 
and correspond to the appearance of the elementary kink itself as a bound state
in the direct and crossed channel, respectively. 

For $q>3$ ($\lambda>1$) a direct channel (positive residue) pole located at
$\theta=2i\kappa$,
\EQ
\kappa=\frac{\pi}{2}\left(1-\frac{1}{\lambda}\right)\,,
\EN
enters the physical strip in the amplitudes $S_2(\theta)$ and $S_3(\theta)$. 
Such a pole must be accordingly associated to a (topologically neutral)
kink-antikink bound state $B$ with mass 
\EQ
m_B=2m\cos\kappa\,\,.
\label{mb}
\EN
The amplitudes $S_1(\theta)$ and $S_3(\theta)$ exhibit the corresponding 
crossed channel (negative residue) pole at $\theta=i\pi-2i\kappa$. 
The amplitudes $S_{KB}(\th)$ and $S_{BB}(\th)$ describing the 
kink-bound  state scattering and the bound state self-interaction are 
determined by the bootstrap equations
\bea
&& S_{KB}(\theta)=(q-2)S_2(\theta-i\kappa)S_1(\theta+i\kappa)+
S_3(\theta-i\kappa)S_3(\theta+i\kappa)\,,\nonumber\\
&& S_{BB}(\theta)=S_{BK}(\theta-i\kappa)S_{BK}(\theta+i\kappa)\,,
\eea
and read
\bea
&& S_{BK}(\theta)=t_{1-\kappa/\pi}(\theta)t_{2/3-\kappa/\pi}(\theta)\,,
\label{bk}\\
&& S_{BB}(\theta)=t_{2/3}(\theta)t_{1-2\kappa/\pi}(\theta)t_{2/3-2\kappa/\pi}
(\theta)\,,
\label{bb}
\eea
in terms of the functions
\EQ
t_a(\theta)=\frac{\tanh\frac{1}{2}(\theta+i\pi a)}{\tanh\frac{1}{2}
(\th-i\pi a)}\,\,.
\EN
The poles located at $\theta=i(\pi-\kappa)$ in $S_{BK}$ and at $\theta=2\pi/3$ 
in $S_{BB}$ are bound state poles corresponding to $K$ and $B$, respectively.

It has been shown \cite{DPT} that the remaining poles in the amplitudes 
$S_{KB}$ and $S_{BB}$ are associated to multi-scattering processes
rather than to new particles. Hence, the elementary kinks and their neutral 
bound state $B$ are the only particles in the spectrum of the field theory
describing the scaling zero-field Potts model below the critical temperature.

\begin{figure}
\centerline{
\includegraphics[width=9cm]{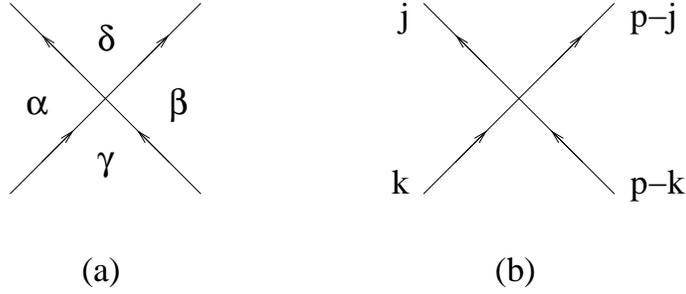}}
\caption{Pictorial representation of the kink scattering amplitudes
$S_{\alpha\beta}^{\gamma\delta}$ (a), and of the high-temperature 
amplitudes $\tilde{S}_{k,p-k}^{p-j,j}$ (b). 
}
\end{figure}

\subsection{Disordered phase}
We now show how the amplitudes (\ref{s0})--(\ref{s3}) determine also the 
scattering theory in the disordered phase.

Above the critical temperature there is a unique ground state and the 
excitations of the scaling limit must be ordinary particles rather than kinks.
Observing that the group $Z_q$ of cyclic permutations is a subgroup of $S_q$,
we can take these particles to have a well defined $Z_q$ charge. The simplest
possibility is then to conjecture a basis of elementary excitations 
$A_k(\theta)$, $k=1,\ldots,q-1$, each carrying $k$ units of $Z_q$ charge. A 
multi-particle state $A_{k_1}(\theta_1)\ldots A_{k_n}(\theta_n)$ will carry
a charge $k_1+\ldots+k_n\,\,(\mbox{mod}\,q)$. The integrable scattering theory
is characterized by the two-particle amplitudes defined through the 
Faddeev-Zamolodchikov algebra (Fig.~3b)
\EQ
A_k(\theta_1)A_{p-k}(\theta_2)=\sum_{j\neq p}\tilde{S}_{k,p-k}^{p-j,j}
(\theta_{12})A_{p-j}(\theta_2)A_j(\theta_1)\,,
\label{fz2}
\EN
where all the particle indices are taken $\mbox{mod}\,q$. The neutral channel
corresponds to $p=q$, while $p=1,\ldots q-1$ yields charged channels. Full
$S_q$ invariance is then recovered requiring that the interaction does not 
distinguish between the charged channels; within the neutral
channel and the charged channel, the need to have well defined crossing 
properties forces to distiguish the case
$j=k$ from the case $j\neq k$. This leaves us with a total number of four 
different amplitudes corresponding to
\bea
&& A_k(\theta_1)A_{p-k}(\theta_2)=\tilde{S}_0(\theta_{12})
\sum_{j\neq k,p}A_j(\theta_2)A_{p-j}(\theta_1)+\tilde{S}_1(\theta_{12})
A_k(\theta_2)A_{p-k}(\th_1)\,,\hspace{.5cm}p\neq q
\nonumber\\
&& A_k(\theta_1)A_{q-k}(\theta_2)=\tilde{S}_2(\theta_{12})
\sum_{j\neq k}A_j(\theta_2)A_{q-j}(\theta_1)+\tilde{S}_3(\theta_{12})
A_k(\theta_2)A_{q-k}(\th_1)\,\,.
\eea
These identifications require that all the particles $A_k$ have the same mass
$\tilde{m}$. 

The unitarity equations
\EQ
\sum_{l\neq p}\,\tilde{S}_{k,p-k}^{p-l,l}(\theta)\,\tilde{S}_{l,p-l}^{p-j,j}
(-\theta)=\delta_k^j\,\delta_{p-k}^{p-j}
\EN
are obtained iterating (\ref{fz2}) and take a form identical to 
(\ref{uni1})--(\ref{uni4}) with the substitution $S_i\to\tilde{S}_i$. The
crossing relations coincide with (\ref{cross1})--(\ref{cross3}) under the same
substitution. 

This correspondence between the scattering theories above and
below the critical temperature extends to the factorization and bootstrap 
equations and simply expresses the fact that the two phases are related by 
duality. The elementary excitations of the disordered phase are particles
$A_k$, $k=1,\ldots,q-1$, with the same mass of the kinks of the low-temperature
phase and whose scattering is expressed in terms of the same amplitudes
which specify the kink $S$-matrix, i.e.
\EQ
\tilde{m}=m\,,\hspace{1.5cm}\tilde{S}_i(\theta)=S_i(\theta)\,,
\hspace{.5cm}i=0,1,2,3\,\,.
\EN
For $3<q\leq 4$, the high-temperature theory contains a neutral bound
state $B$ with the mass (\ref{mb}) and whose scattering is specified by 
amplitudes $S_{BA_k}$ and $S_{BB}$ coinciding with (\ref{bk}) and (\ref{bb}). 

Duality allows to compute correlation functions above and below $T_c$ within
the form factor approach which relies on the knowledge of the $S$-matrix. 
This was done in \cite{DC} using the kink $S$-matrix. It can be checked 
that the same results are obtained through the high-temperature scattering 
theory.

\section{Particle spectrum in weak magnetic field}
The action (\ref{scaling}) describes the renormalization group trajectories
flowing out of the fixed point located at the origin of the $\tau$--$h$ plane.
Such trajectories can be labelled by the dimensionless parameters
\EQ
\eta_\pm=\frac{\tau}{(\pm h)^{(d-X_\varepsilon)/(d-X_\sigma)}}\,,
\label{eta}
\EN
where the upper and lower signs are used for $h>0$ and $h<0$, 
respectively, in such a way that $\eta_+$ parameterizes the trajectories in
the upper half--plane and $\eta_-$ those in the lower half--plane; the two
trajectories at $h=0$ correspond to $\eta_\pm=+\infty$ and 
$\eta_\pm=-\infty$ (see Fig.~4).

\begin{figure}
\centerline{
\includegraphics[width=7cm]{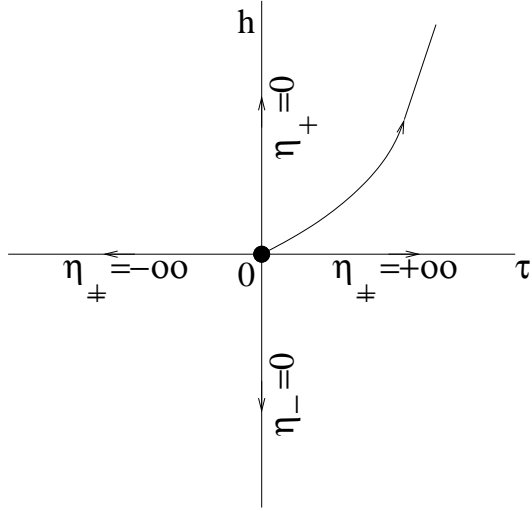}}
\caption{Some renormalization group trajectories associated to the action
(\ref{scaling}).
}
\end{figure}

In this section we discuss the evolution of the mass spectrum in $d=2$
starting from the case of small magnetic field for which perturbation theory
around integrable quantum field theories \cite{nonint} can be used.

\subsection{Weak field above critical temperature}
For small $h$ at $\tau>0$ (i.e. for $\eta_\pm\to+\infty$) the
corrections to the mass spectrum are determined by the matrix elements of 
the magnetic operator $\sigma_q(x)$ on the asymptotic states of the unperturbed
($\eta_\pm=+\infty$) theory. At leading order, the correction to the 
mass matrix is \cite{nonint} (Fig.~5a)
\EQ
(\delta m^2)_{j,k}\simeq -2h\,\langle A_j(0)|\sigma_q(0)|A_k(0)\rangle\,,
\hspace{1cm}j,k=1,\ldots,q-1
\label{dm}
\EN
(here and below the matrix element on the particles $A_k$ are intended at 
$h=0$). 

Let us introduce the operators
\EQ
\tau_k(x)=\sum_{\alpha=1}^q \omega_q^{-\alpha k}\,\sigma_\alpha(x)\,,
\hspace{1.5cm}k=1,\ldots,q-1
\EN
where $\omega_n\equiv \exp(2i\pi/n)$. The generator $\Omega_q$ of cyclic 
permutations acts as
\EQ
\Omega_q\,\sigma_\alpha(x)=\sigma_{\alpha+1\,(\mbox{mod}\,q)}(x)\,,
\hspace{1.5cm}\Omega_q\,\tau_k(x)=\omega_q^k\,\tau_k(x)\,,
\EN
showing that $\tau_k$ carries $k$ units of $Z_q$ charge and can be taken as 
interpolating operator of particle $A_k$ at $h=0$. Using 
\EQ
\sum_{j=1}^n\omega_n^{\pm jk}=n\,\delta_{k,n}
\EN
and (\ref{constraint}), one also have
\EQ
\sigma_q(x)=\frac{1}{q}\,\sum_{k=1}^{q-1}\tau_k(x)\,\,.
\label{sigmaq}
\EN
Denoting 
\EQ
M_{j,k}^\Psi=\langle A_j(0)|\Psi(0)|A_k(0)\rangle\,,
\EN
and taking into account conservation of $Z_q$ charge as well as full $S_q$
symmetry at $h=0$, we have
\EQ
M_{j,k}^{\tau_l}=f_q\,\delta_{j,k+l\,(\mbox{mod}\,q)}
\EN
and
\EQ
M_{j,k}^{\sigma_q}=\frac{1}{q}\,\sum_{l=1}^{q-1}M_{j,k}^{\tau_l}=\frac{f_q}{q}
\,(1-\delta_{j,k})
\EN
for the matrix which determines the mass corrections (\ref{dm}). 
Diagonalization of this matrix gives for the spectrum at $\eta_\pm\to+\infty$
a particle
\EQ
A_0=\frac{1}{\sqrt{q-1}}\,\sum_{k=1}^{q-1}\,A_k
\label{a0}
\EN
with square mass 
\EQ
m_0^2\simeq m^2-2f_q\,\frac{q-2}{q}\,h\,,
\label{m0}
\EN 
and a degenerate multiplet
\EQ
A_k'=\frac{1}{\sqrt{2}}\,(A_k-A_{q-1})\,,\hspace{1.5cm}k=1,\ldots,q-2
\label{ak'}
\EN
with square mass
\EQ
m'^2\simeq m^2+2\,\frac{f_q}{q}\,h\,\,.
\label{m'}
\EN
Comparison with (\ref{sigmaq}) shows that $A_0$ is interpolated by $\sigma_q$ 
and is a singlet of the $S_{q-1}$ symmetry surviving at $h\neq 0$; suitable 
linear combinations of the $A_k'$ yield a multiplet in which each of the $q-2$ 
particles carries a definite (non-zero) $Z_{q-1}$ charge.

\begin{figure}
\centerline{
\includegraphics[width=10cm]{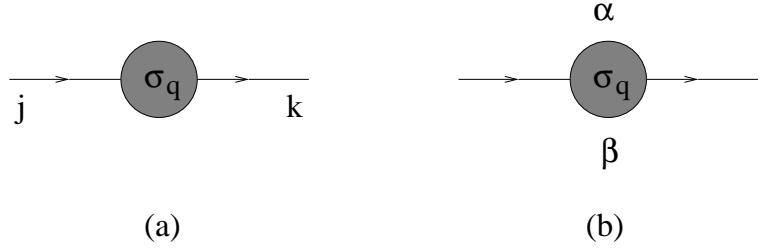}}
\caption{First order mass corrections in weak magnetic field above critical 
temperature (a), and below (b).
}
\end{figure}

For $3<q\leq 4$ the theory also contains the additional $S_{q-1}$--singlet
$B$ with a mass correction with respect to (\ref{mb}) given by
\EQ
\delta m_B^2\simeq -2h\,\langle B(0)|\sigma_q(0)|B(0)\rangle\,\,.
\label{dmb}
\EN
This matrix element, as well as $f_q$ above, can be obtained from the form 
factor results of \cite{DC} (for $q=4$ see also \cite{DG}). Notice that the
first order correction to the mass of $A_0$ vanishes for $q=2$; the leading 
correction in this case is of order $h^2$ and was computed in \cite{FZ2}.

\subsection{Weak field below critical temperature}
A small magnetic field acting on the sites with color $q$ affects the ground 
state degeneracy at $\tau<0$. Denoting by $|0_\alpha\rangle$, $\alpha=1,\ldots,
q$ the ferromagnetic vacua, $S_q$ symmetry gives at $\eta_\pm=-\infty$ 
\EQ
\langle\sigma_\gamma\rangle_\alpha\equiv
\langle 0_\alpha|\sigma_\gamma(x)|0_\alpha\rangle=\frac{v}{q-1}\,(q\,\delta_{
\gamma,\alpha}-1)\,,
\label{vevs}
\EN
with $v$ positive. At first order in $h$, the energy density 
difference between a vacuum 
$|0_{\alpha\neq q}\rangle$ and $|0_q\rangle$ is then
\EQ
\Delta{\cal E}=\delta{\cal E}_\alpha-\delta{\cal E}_q\simeq-h\,
(\langle\sigma_q\rangle_\alpha-\langle\sigma_q\rangle_q)=\frac{v\,q}{q-1}\,h\,,
\label{deltae}
\EN
so that $|0_q\rangle$ is the unique true vacuum at $h>0$, and the unique false 
vacuum at $h<0$. 

Since no finite--energy topological excitation can begin or end on a false 
vacuum, 
the space of asymptotic states of the theory does not contain kinks for $h>0$. 
For $\eta_-$ large and negative, instead, the kinks $K_{\alpha\beta}$ with 
$\alpha,\beta\neq q$ survive as the elementary excitations of the theory.
The first order correction to their mass is (Fig~5b)
\EQ
\delta m^2_{K_{\alpha\beta}}\simeq -2h\,\langle K_{\alpha\beta}(0)|\sigma_q(0)|
K_{\beta\alpha}(0)\rangle_{conn}=-2h\,F^{\sigma_q}_{\alpha\beta\alpha}(i\pi)\,,
\label{dmk}
\EN
where
\EQ
F^{\sigma_q}_{\alpha\beta\alpha}(\theta_1-\theta_2)=\langle 0_\alpha|
\sigma_q(0)|K_{\alpha\beta}(\theta_1)K_{\beta\alpha}(\theta_2)\rangle\,,
\hspace{1cm}\alpha\neq\beta
\EN
is the two-kink form factor at $\eta_\pm=-\infty$ \cite{DC}. In general this 
function has an annihilation pole whose residue 
\EQ
-i\,\mbox{Res}_{\theta=i\pi}\,F^{\sigma_q}_{\alpha\beta\alpha}(\theta)=
\langle\sigma_q\rangle_\alpha-\langle\sigma_q\rangle_\beta=
\frac{v\,q}{q-1}\,(\delta_{\alpha,q}-\delta_{\beta,q})
\EN
vanishes precisely when both $\alpha$ and $\beta$ differ from $q$, giving
a finite mass correction\footnote{In this case (\ref{dmk}) is finite 
irrespectively of the sign of $h$. We should recall, however, that for $h>0$
we are computing a mass gap above a false vacuum whose energy separation 
from the true vacuum in a system of spatial size $L$ is $\Delta{\cal E}\,L$.
Hence, all the single-kink states decouple in the infinite system with a 
positive, however 
small, magnetic field.} (\ref{dmk}) (which does not depend on $\alpha,\beta
=1,\ldots,q-1$). The divergence of (\ref{dmk}) when $\alpha$ or 
$\beta$ equal $q$ simply reflects the decoupling of a kink interpolating 
between vacua which are no longer degenerate. 

\vspace{.3cm}
Although for $h$ positive $|0_q\rangle$ is the only true vacuum and 
no single-kink state survives, $n$-kink states beginning and ending on the 
true vacuum, i.e. $K_{q\alpha_1}(\theta_1)K_{\alpha_1\alpha_2}(\theta_2)\ldots
K_{\alpha_{n-1}q}(\theta_n)$, do not decouple. Here we consider all the 
intermediate vacua to be false, since insertion of a true intermediate 
vacuum would simply amount to breaking the sequence into two states of the 
first type. Then we have $n\leq q$.

\begin{figure}
\centerline{
\includegraphics[width=12cm]{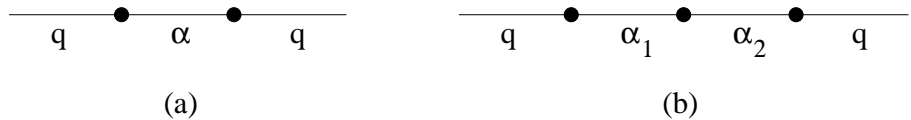}}
\caption{Spatial kink configurations which are confined into mesons (a) and
baryons (b) by a positive magnetic field.
}
\end{figure}

In the non-relativistic limit
valid for small rapidities, the total energy of such a configuration consist
of the rest mass term $n\,m$, the kinetic term, a contribution coming from 
kink interaction which decays exponentially with interkink distance, and the
false vacuum contribution
\EQ
V(x_1,\ldots,x_n)=\Delta{\cal E}\,\sum_{i=1}^{n-1}(x_{i+1}-x_i)=
(x_n-x_1)\,\Delta{\cal E}\,,
\label{vconf}
\EN
where $x_1<x_2<\ldots<x_n$ are the spatial positions of the kinks. The 
positive linear potential
(\ref{vconf}) confines the $n$ kinks into a finite spatial interval and 
prevents the observation of isolated kinks. In this sense the kinks of the
$\eta_\pm=-\infty$ theory play the role of ``quarks'' at $h>0$. The asymptotic
particles are instead the topologically neutral bound states produced by the
confinement of the $n$-kink state. When $\eta_+\to-\infty$ the confining 
potential is extremely shallow and the kinks are very loosely bound; the 
average interkink distance is large and kink interaction is negligible in first
approximation. The 
$n$-kink bound states form an infinite tower of levels which are dense 
above the value $n\,m$ as $\eta_+\to-\infty$. It is natural to call ``mesons''
the $n=2$ (kink-antikink) bound states (Fig.~6a), and ``baryons'' the bound 
states with $n=3$ (Fig.~6b). Recalling the conditions $n\leq q$ and 
$q\leq q_c=4$, we have that the mesons can occur for $q=2,3,4$, and the baryons
for $q=3,4$;  tetraquark confined states are allowed for $q=4$.

These particles organize themselves into multiplets of the residual $S_{q-1}$
symmetry. The number of different $n$-kink sequences coincides with the 
possible ways of coloring the intermediate vacua. The $q-1$ mesonic 
sequences can be combined into the states
\EQ
\pi_k^{(j)}(0)\sim\sum_{\alpha=1}^{q-1}\,\omega_{q-1}^{-k\alpha}\,
K_{q\alpha}(\theta)K_{\alpha q}(-\theta)\,,\hspace{1.5cm}k=0,1,\ldots,q-2
\label{pi}
\EN
with $j=1,2,\ldots$ labelling in order of increasing energy the levels 
originated by the confinement of the kink-antikink superposition. Since
\EQ
\Omega_{q-1}\,\pi_k^{(j)}=\omega^k_{q-1}\,\pi_k^{(j)}\,,
\EN 
the mesons (\ref{pi}) are eigenstates under cyclic permutations of the first 
$q-1$ colors with $Z_{q-1}$ charge $k$. In the non-relativistic limit valid
for the lowest levels in weak field, we can think of the kink and antikink 
inside a meson as experiencing elastic reflection on the walls of the confining
potential and elastic scattering among themselves. As the effect of the 
latter, the intermediate vacuum can either remain unchanged with a probability
amplitude $\Sigma_3(2\theta)$, or switch to a different color with a
probability amplitude $\Sigma_2(2\theta)$ which, by $S_{q-1}$-invariance, does
not depend on the new color\footnote{At $\eta_\pm=-\infty$ the scattering 
amplitudes $\Sigma_2$ and $\Sigma_3$ coincide with (\ref{s2}) and (\ref{s3}), 
respectively. Corrections in weak field are determined by the matrix elements
of the magnetic operator $\sigma_q$ \cite{nonint}.}. The superpositions in the 
right hand side of (\ref{pi}) are eigenstates of the $S$-matrix with scattering
amplitudes 
\EQ
\Sigma_3(2\theta)+[(q-1)\delta_{k,0}-1]\Sigma_2(2\theta)\,\,.
\label{sigma23}
\EN
Hence, the quark interaction is different for the neutral mesons ($k=0$)
and for the charged mesons ($k\neq 0$). For fixed $j$ this will lead to a 
siglet $\pi_0^{(j)}$ and a multiplet $\pi_1^{(j)},
\ldots,\pi_{q-2}^{(j)}$ with energies which differ very slightly in weak 
field.

\begin{figure}
\centerline{
\includegraphics[width=4.5cm]{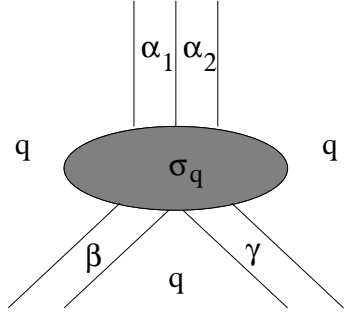}}
\caption{Seven-kink matrix element responsible for the decay of baryons 
above threshold into two mesons in positive magnetic field.
}
\end{figure}

Only a finite number of these confined states are stable. For $q=2,3$ all 
those states with energy larger than twice the mass of the lightest mesons 
(i.e. larger than a value close to $4m$ for weak field) lie in the continuum 
and can decay. This means that for $\eta_+$ sufficiently large and negative we
certainly have stable mesons in the energy interval $(2m,4m)$ for $q=2,3$, and 
stable baryons in the energy interval $(3m,4m)$ for $q=3$. At $q=4$ the decay
thresholds are lowered by the presence of the neutral particle $B$ with 
unperturbed\footnote{The first order correction to $m_B^2$ coincides with
(\ref{dmb}).} mass $m_B=\sqrt{3}m$. Hence the stability intervals in weak field
are $(2m,2\sqrt{3}m)$ for the neutral mesons, $(2m,(2+\sqrt{3})m)$ for the 
charged mesons, $(3m,2\sqrt{3}m)$ for the neutral baryons and 
$(3m,(2+\sqrt{3})m)$ for the charged baryons; as for the tetraquark states, 
they all lie in the continuum and are expected to be unstable.

The number of $3$-kink sequences giving rise to baryons is $(q-1)(q-2)$. 
For $q=3$ we then have the two series of baryons
\EQ
p_\pm^{(j)}(\theta)\sim K_{31}(\theta_1)K_{12}(\theta_2)K_{23}(\theta_3)\pm
K_{32}(\theta_1)K_{21}(\theta_2)K_{13}(\theta_3)\,,
\EN
with even or odd parity with respect the residual $Z_2$ symmetry which 
interchanges the colors 1 and 2. For $q=4$ the residual $S_3$ symmetry
can be seen as the product of the group $Z_3$ of cyclic permutations times
the topological charge conjugation $C_T$ which transforms the kink 
$K_{\alpha\beta}$, $\alpha,\beta\neq 4$, into its antikink $K_{\beta\alpha}$. 
Then the six series of baryonic states
\bea
p_{k,\pm}^{(j)}(\theta)\sim\sum_{\alpha=1}^3\omega_3^{-k\alpha}&&\left[ 
K_{4,\alpha}(\theta_1)K_{\alpha,\alpha+1\,(\mbox{mod}\,3)}(\theta_2)
K_{\alpha+1\,(\mbox{mod}\,3),4}(\theta_3)\,\,\pm\right.\nonumber \\
&&\left.\,\,
K_{4,\alpha+1\,(\mbox{mod}\,3)}(\theta_1)K_{\alpha+1\,(\mbox{mod}\,3),\alpha}
(\theta_2)K_{\alpha,4}(\theta_3)\right]\,,\hspace{.6cm}k=0,1,2
\eea
are $S_3$--eigenstates:
\EQ
\Omega_3\,p_{k,\pm}^{(j)}=\omega_3^k\,p_{k,\pm}^{(j)}\,,\hspace{1.5cm}
C_T\,p_{k,\pm}^{(j)}=\pm\,p_{k,\pm}^{(j)}\,\,.
\EN
For a given $j$ they give rise to two singlets ($k=0$) and two doublets
($k\neq 0$) with opposite $C_T$--parity.

\vspace{.3cm}
As already said, a small negative magnetic field confines only the kinks 
$K_{\alpha\beta}$ with $\alpha$ or $\beta$ equal $q$. Since now the false 
vacuum is unique, the 
only confined states are mesons of type $K_{\alpha q}K_{q\alpha}$. For $q>2$, 
however, these can decay into two asymptotic kinks through expansion of bubbles
of true vacuum into the false vacuum.

\section{Evolution of the spectrum with temperature and magnetic field}

\subsection{Positive magnetic field}
We discussed in the previous section how the particle spectrum of the scaling
two-dimensional $q$-state Potts model changes when a small magnetic field 
acting on the sites with color $q$ is switched on. For positive field, below 
critical temperature we have a complete confinement of kinks and the generation
of a dense spectrum of mesons and (for $q>2$) baryons; the number of such
particles which are stable tends to infinity as $\eta_+\to-\infty$. Above 
critical temperature, on the other hand, the effect of the magnetic field is 
much less dramatic, simply amounting to a partial removal of the degeneracy of 
the mass spectrum of the $h=0$ particles. 

It must be possible to interpolate continuously between these two limiting 
cases following the evolution of the spectrum as $\eta_+$ grows from $-\infty$ 
to $+\infty$ on the plane of Fig.~4. The 
simplest scenario is that, as $\eta_+$ increases, more and more mesons and 
baryons cross the decay thresholds and become unstable. This process of 
depletion of the spectrum of stable excitations would continue
until the only stable particles surviving as $\eta_+\to+\infty$ are those of 
the high-temperature theory in zero field. 

We expect that, during the evolution of the spectrum as a function of the 
parameters $\eta_\pm$, energy levels corresponding to states originating from
the confinement of a same number of kinks do not cross for $h\neq 0$. Indeed, 
the additional degeneracy at a crossing point of this nature should normally be
related to a symmetry enhancement. In the 
field theory (\ref{scaling}) the only symmetry enhancement (from $S_{q-1}$ to 
$S_q$) occurs at $h=0$. If we add to this that the baryons should decay more
easily than the lightest mesons, it seems natural to expect that the particles 
surviving in the limit $\eta_+\to+\infty$ should be identified with the 
lightest among the mesons produced by kink confinement at 
$\eta_+$ very large and negative. In particular, the lightest meson multiplet 
$\pi_{1}^{(1)},\ldots,\pi_{q-2}^{(1)}$ should evolve for increasing $\eta_+$ 
into the multiplet (\ref{ak'}). As for the particle which evolves into the 
singlet (\ref{a0}), it should be identified with $\pi_{0}^{(1)}$ for $q=2,3$. 
At $q=4$ the zero-field spectrum also includes the neutral particle $B$ with 
mass $\sqrt{3}m$, which is the lightest particle for $\eta_+\to-\infty$. If one
assumes that it does not cross mesonic levels, then it should evolve into the 
neutral particle (\ref{a0}) for $\eta_+\to+\infty$; then the meson 
$\pi_{0}^{(1)}$ could evolve into the particle with mass $\sqrt{3}m$ in the 
same limit. 

According 
to this scenario, all the mesons (\ref{pi}) with $j>1$ and all the baryons
must have become unstable by the time $\eta_+$ approaches $+\infty$. For each
of this particles there should exist a finite ``critical'' value of $\eta_+$
for which they reach the lowest decay threshold compatible with their charge, 
and above which they become unstable. Such critical value is expected to 
decrease as the mass of the particle increases, so that the ``critical 
trajectories'' accumulate in the limit $\eta_+\to-\infty$. Notice that, due
to the non-locality of the magnetic operator with respect to the kinks, the 
magnetic term of the action has infinitely many matrix elements on kinks which
are non-zero in zero-field. In particular, the vertex responsible for the 
decay of the baryons which reach the two-meson threshold is shown in (Fig.~7).

Also in view of considerations to be made below about the spectrum evolution 
at $h<0$, we expect the quantity $f_q$ determining the first order mass
corrections (\ref{m0}) and (\ref{m'}) to be positive\footnote{This
quantity can be exactly computed within the integrable field theory at $h=0$
(see \cite{DC}).}. Together with the previous identifications, this leads to 
the expectation 
that for any $h>0$ the different quark interactions inside the neutral and the 
charged mesons (see (\ref{sigma23})) induce a mass splitting between 
$\pi_{1}^{(1)},\ldots,\pi_{q-2}^{(1)}$ and $\pi_{0}^{(1)}$ which is positive
for $q=3$. If the previous speculations about $q=4$ should turn out to be 
correct, the mass splitting would be negative in this case.

When specialized to $q=2$ this scenario coincides with that originally proposed
for the scaling Ising model by McCoy and Wu \cite{McW}. In this case the
pattern is simplified by the absence of baryons and charged mesons, as well as
by the absence of interaction at $h=0$. Moreover, integrability at $\tau=0$ 
\cite{Taniguchi} also allows analytic investigation for strong magnetic field. 
Several studies, both analytic and numerical, have confirmed the McCoy-Wu
scenario and provide us with a detailed 
description of the mass spectrum of Ising field theory in the full $\tau$--$h$ 
plane \cite{McW,Taniguchi,nonint,FZ1,FZ2,decay,Rutkevich,FZ3,GR,CGR} (see also
\cite{review} for an introductory review of Ising field theory).

\begin{figure}
\centerline{
\includegraphics[width=12cm]{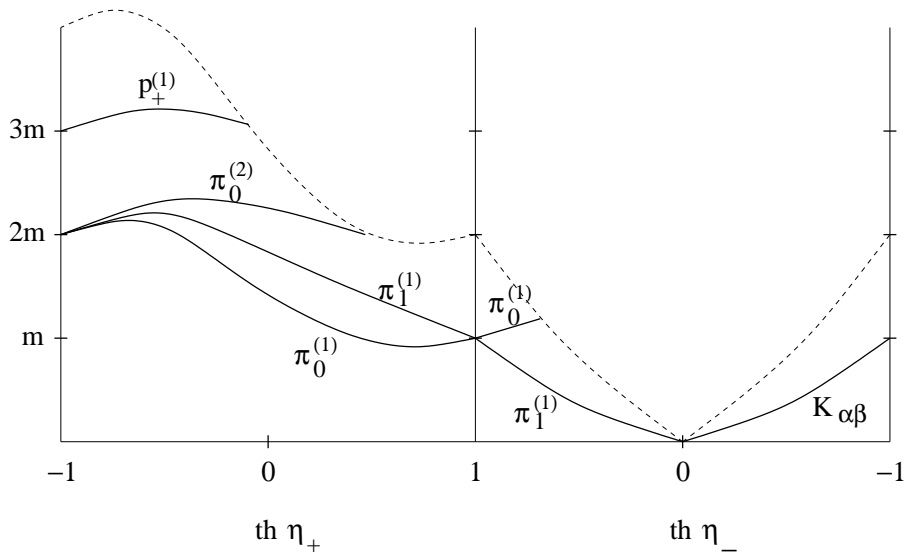}}
\caption{Conjectured qualitative evolution of the mass spectrum for $q=3$.
The left-hand half corresponds to the evolution in $\eta_+$, the 
right-hand half to the evolution in $\eta_-$. The values $\eta_\pm=+\infty$, as
well as the values $\eta_\pm=-\infty$, describe the same renormalization group 
trajectory. Only three lightest mesons and the lightest neutral baryon are 
shown for positive magnetic field. The 
dashed line is the lowest decay threshold (twice the mass of the lightest 
particle). Particles whose mass reaches the threshold become unstable. 
Deconfined kinks exist for negative $\eta_-$ only.
}
\end{figure}

\subsection{Negative magnetic field}
We now discuss the evolution of the spectrum as a function of $\eta_-$ for 
$q=3,4$ (for $q=2$ the spectrum does not depend on the sign of the magnetic 
field). We saw in the previous section that the elementary excitations for 
$\eta_-\to-\infty$ are the kinks $K_{\alpha\beta}$ interpolating between the 
$q-1$ residual vacua. For $\eta_-\to+\infty$, on the other hand, due to the 
negative sign of $h$ in (\ref{m0}) and (\ref{m'}), the lightest excitations are
the charged particles that in the previous subsection we identified with 
the mesons $\pi_{1}^{(1)},\ldots,\pi_{q-2}^{(1)}$. Passing from these
topologically neutral particles to the topologically charged kinks requires
a phase transition at an intermediate value of $\eta_-$. We already know that, 
for $q=3,4$, a second order phase transition takes place at $\eta_-=0$
\footnote{An analogous transition from ordinary particles to kinks, involving 
the spontaneous breaking of the symmetry $S_q$ rather than $S_{q-1}$, takes 
place when going through $\tau=0$ at $h=0$.}. The mass
of the kinks decreases as $\eta_-$ goes from $-\infty$ to zero, and vanishes 
at $\eta_-=0$. Similarly, the mass of the mesons $\pi_{1}^{(1)},\ldots,
\pi_{q-2}^{(1)}$ decreases as $\eta_-$ goes from $+\infty$ to zero, and 
vanishes at $\eta_-=0$. Deconfined kinks exist only
for $\eta_-\in[-\infty,0)$; for $\eta_-\in(0,+\infty]$ and for all positive 
values of $h$ the spectrum can be seen as consisting of kink bound states.
The trajectory $\eta_-=0$ is a massless flow between the $S_q$-invariant
ultraviolet fixed point at $h=0$ and the $S_{q-1}$-invariant infrared fixed 
point at $h=-\infty$.

The ratio between the mass of the neutral meson $\pi_{0}^{(1)}$ and that of the
charged mesons $\pi_{1}^{(1)},\ldots,\pi_{q-2}^{(1)}$, which is $1$ at 
$\eta_-=+\infty$, increases as $\eta_-$ decreases until it reaches the value
$2$ for some positive value of $\eta_-$. Beyond this point the opening of the
decay channel $\pi_{0}^{(1)}\to\pi_{1}^{(1)}\pi_{q-2}^{(1)}$ makes the neutral
meson unstable. The second neutral particle present at $q=4$ becomes unstable
in the same way. We then expect that the charged mesons are the only stable
particles surviving for $\eta_-$ sufficiently small and positive. Similarly for
the kinks at $\eta_-$ sufficiently small and negative. 

The conjectured evolution of the mass spectrum with the parameters $\eta_+$ and
$\eta_-$ is illustrated in Fig.~8 for $q=3$.

\section{Conclusion}
In this paper we considered the field theory describing the scaling limit of 
the two-dimensional $q$-state Potts model in a magnetic field acting on one of 
the $q$ colors. This field breaks the $S_q$ symmetry of color permutations 
down to $S_{q-1}$ and, for $2<q\leq q_c=4$ allows for an extended region in 
the plane of temperature and magnetic field in which the quarks (kinks 
interpolating between degenerate vacua) are deconfined. If the analysis is 
extended to a more general magnetic term $\sum_{\alpha_1}^{q-1} h_\alpha
\sigma_\alpha$, a renormalization group trajectory corresponding to generic 
values of the components $h_\alpha$ will possess no internal symmetry and will
contain only particles made of confined quarks with unequal
masses. Of course, suitable relations among the magnetic parameters identify 
regions in parameter space with a residual $S_{q-1}$ or (for $q=4$) $S_{q-2}$
symmetry, which in turn contain phases with deconfined quarks. 

We saw how form factor perturbation theory allows to compute mass corrections
in weak field above critical temperature, and below critical temperature for 
the quarks in the deconfined phase. As for the mass spectrum of the particles 
originating from confinement, the Bethe-Salpeter approach proved to be 
remarkably
effective for the Ising mesons \cite{FZ1,FZ3}, but needs to be generalized to
the case of quarks which interact already in absence of field in order to deal
with $q\neq 2$. Hopefully, it will become possible to study also the
baryonic spectrum along similar lines. 

Numerical methods will be eventually needed for a quantitative study of the 
mass spectrum in strong field. Particularly promising in this respect seems the
truncated conformal space approach \cite{trunc}, which amounts to the numerical
diagonalization of the Hamiltonian on a finite-dimensional subspace of the 
conformal basis of states of the ultraviolet fixed point. This approach has
been successfully used for the Ising case\footnote{In \cite{FZ1,FZ3} the free 
nature of the zero-field Ising model is exploited to diagonalize the
Hamiltonian on a truncated basis of massive fermionic states.}  
\cite{nonint,FZ1,FZ3,PT} and can be used also for $q\neq 2$.

\vspace{1cm}
{\bf Acknowledgments.}\hspace{.3cm} The work of G.D. is partially supported by 
the ESF grant INSTANS and by the MUR project ``Quantum field theory and 
statistical mechanics in low dimensions''.


\newpage
\begin{thebibliography}{99}

\bibitem{CJS} S. Coleman, R. Jackiw and L. Susskind, Ann. Phys. 93 (1975) 267.
\bibitem{Coleman} S. Coleman, Ann. Phys. 101 (1976) 239.
\bibitem{McW} B.M. McCoy and T.T. Wu, Phys. Rev. D 18 (1978) 1259.
\bibitem{dsg} G. Delfino and G. Mussardo, Nucl. Phys. B 516 (1998) 675.
\bibitem{Potts} R.B. Potts, Proc. Cambridge Phil. Soc. 48 (1952) 106.
\bibitem{Wu} F.Y. Wu, Rev. Mod. Phys. 54 (1982) 235.
\bibitem{Baxter} R.J. Baxter, Exactly solved models of statistical
mechanics, Academic Press, London (1982).
\bibitem{DG} G. Delfino and P. Grinza, Nucl. Phys. B 682 (2004) 521.
\bibitem{CZ} L. Chim and A.B. Zamolodchikov, Int. J. Mod. Phys. A 7 (1992) 
5317.
\bibitem{nonint} G. Delfino, G. Mussardo and P. Simonetti, Nucl. Phys. B 473
(1996) 469.
\bibitem{DC} G. Delfino and J. Cardy, Nucl. Phys. B 519 (1998) 551.
\bibitem{KF} P.W. Kasteleyn and E.M. Fortuin, J. Phys. Soc. Jpn. Suppl.
26 (1969), 11; Physica 57 (1972) 536.
\bibitem{DF} Vl.S. Dotsenko and V.A. Fateev, Nucl. Phys. B 240 (1984) 312.
\bibitem{BPZ} A.A. Belavin, A.M. Polyakov and A.B. Zamolodchikov,
Nucl. Phys. B 241 (1984) 333.
\bibitem{Nienhuis} B. Nienhuis, J. Stat. Phys. 34 (1984) 781.
\bibitem{NBRS} B. Nienhuis, A. Berker, E. Riedel and M. Shick, Phys. Rev. Lett.
43 (1979) 737.
\bibitem{CNS} J. Cardy, M. Nauenberg and D. Scalapino, Phys. Rev. B 22 (1980)
2560.
\bibitem{dilute} G. Delfino, Nucl. Phys. B 554 (1999) 537.
\bibitem{q4} G. Delfino and J. Cardy, Phys. Lett. B 483 (2000) 303.
\bibitem{ZZ} A.B. Zamolodchikov and Al.B. Zamolodchikov, Ann. Phys. 120 (1979) 
253.
\bibitem{Smirnov12} F.A. Smirnov, Int. J. Mod. Phys. A 6 (1991) 1407.
\bibitem{FR} P. Fendley and N. Read, hep-th/0207176.
\bibitem{DPT} P. Dorey, A. Pocklington and R. Tateo, Nucl. Phys. B 661 (2003)
425.
\bibitem{FZ2} P. Fonseca and A.B. Zamolodchikov, hep-th/0309228.
\bibitem{Taniguchi} A.B. Zamolodchikov, Advanced Studies in Pure
Mathematics 19 (1989) 641; Int. J. Mod. Phys. A 3 (1988) 743.
\bibitem{FZ1} P. Fonseca and A.B. Zamolodchikov, J. Stat. Phys. 110 (2003) 527.
\bibitem{decay} G. Delfino, P. Grinza and G. Mussardo, Nucl. Phys. B 737 (2006)
291.
\bibitem{Rutkevich} S.B. Rutkevich, Phys. Rev. Lett. 95 (2005) 250601.
\bibitem{FZ3} P. Fonseca and A.B. Zamolodchikov, hep-th/0612304.
\bibitem{GR} P. Grinza and A. Rago, Nucl. Phys. B 651 (2003) 387.
\bibitem{CGR} M. Caselle, P. Grinza and A. Rago, hep-lat/0408044.
\bibitem{review} G. Delfino, J. Phys. A 37 (2004) R45.
\bibitem{trunc} V.P. Yurov and Al.B. Zamolodchikov, Int. J. Mod. Phys. A6 
(1991) 4557.
\bibitem{PT}B. Pozgay and G. Takacs, hep-th/0604022.

\end{thebibliography}
\end{document}